\documentclass[times, preprint, 3p]{elsarticle}

\usepackage{hyperref}
\hypersetup{colorlinks=true, breaklinks=true}
\usepackage[usenames,dvipsnames]{color}
\usepackage{soul}
\usepackage{multirow}
\usepackage{nicefrac}
\usepackage{amsmath}
\usepackage{amssymb}
\usepackage{microtype}

\journal{Nucl.\ Instrum.\ and Meth.\ in Phys.\ Res.\ A}

\newcommand{\nuc}[2]{${}^{#1}\mathrm{#2}$}

\newcommand{\labr}{LaBr$_3$}
\newcommand{\cebr}{CeBr$_3$}
\newcommand{\figref}[1]{Figure~\ref{#1}}
\newcommand{\edit}[1]{{#1}}

\begin{document}

\begin{frontmatter}

\title{\edit{First In-Beam Demonstration of} a hybrid {LaBr$_3$}/{CeBr$_3$}/{BGO} array to measure radiative capture resonance energies in an extended gas target using a novel time of flight technique}

\author[SMU]{G.~Christian\corref{mycorrespondingauthor}}
\ead{Greg.Christian@smu.ca}
\cortext[mycorrespondingauthor]{Corresponding author}
\author[TRIUMF]{D. Hutcheon}

\author[McMaster]{I. Casandjian}
\author[NPL,Surrey]{S.\ M.\ Collins}
\author[SMU]{A.\ C.\ Edwin}
\author[SMU]{E. Desmarais}
\author[Mines]{U. Greife}
\author[TRIUMF]{A. Katrusiak}
\author[TRIUMF]{A. Lennarz}
\author[TRIUMF,UVic]{M. Loria}
\author[UPC]{S. Moll\'o}
\author[SMU]{J. O'Connell}
\author[Surrey,Bucharest]{S. Pascu}
\author[UPC]{L. Pedro-Botet}
\author[Surrey]{Zs. Podoly\'ak}
\author[Surrey]{B.\ J. Reed\fnref{reedpresent}}
\fntext[reedpresent]{Present address: TRIUMF, 4004 Wesbrook Mall, Vancouver, BC B6T~2A3, Canada and Department of Astronomy \& Physics, Saint Mary's University, Halifax, NS B3H~3C3, Canada}
\author[NPL,Surrey]{P.\ H. Regan}
\author[TRIUMF,UVic]{C. Ruiz}
\author[NPL]{R. Shearman}
\author[TRIUMF]{S. Upadhyayula\fnref{tejapresent}}
\fntext[tejapresent]{Present address: Lawrence Livermore National Laboratory, Livermore, CA 94550 USA}
\author[TRIUMF]{L. Wagner\fnref{louispresent}}
\fntext[louispresent]{Present address: Facility for Rare Isotope Beams, Michigan State University, East Lansing, MI 48824 USA}
\author[Surrey]{M. Williams}

\address[SMU]{Department of Astronomy \& Physics, Saint Mary's University, Halifax, NS B3H~3C3, Canada}
\address[TRIUMF]{TRIUMF, 4004 Wesbrook Mall, Vancouver, BC B6T~2A3, Canada}
\address[McMaster]{Department of Physics \& Astronomy, McMaster University, Hamilton, ON L8S~4M1, Canada}
\address[NPL]{National Physical Laboratory, Teddington, TW11~0LW, UK}
\address[Surrey]{School of Mathematics \& Physics, University of Surrey, GU2~7XH, UK}
\address[Mines]{Department of Physics, Colorado School of Mines, Golden, Colorado 80401, USA}
\address[UVic]{Department of Physics \& Astronomy, University of Victoria, Victoria, BC V8W~3P6, Canada}
\address[Bucharest]{National Institute for Physics and Nuclear Engineering, R-77125, Bucharest-Magurele, Romania}
\address[UPC]{Departament de F\'isica, Universitat Polit\`ecnica de Catalunya, Barcelona, Spain}

\begin{abstract}
We have deployed a new hybrid array of \labr{}, \cebr{}, and BGO scintillators for detecting $\gamma$~rays at the DRAGON recoil separator at TRIUMF. The array was developed to improve $\gamma$-ray timing resolution over the existing BGO array. This allows the average position of resonant capture in an extended gas target to be determined with ${\sim} 15$~mm precision or better, even with five or fewer detected capture events. This, in turn, allows determination of resonant capture energies with statistical uncertainties below ${\sim} 1\%$.
Here we report the results of a \edit{first in-beam demonstration of the array,} measuring the $E_{cm} = 0.4906(3)$~MeV resonance in the $^{23}\mathrm{Na}(p,\gamma){}^{24}\mathrm{Mg}$ reaction, focusing on the timing properties of the array and its anticipated performance in future experiments with radioactive beams.
\end{abstract}


\end{frontmatter}


\section{Introduction}
\label{sec:Intro}

Radiative proton and alpha capture to narrow, isolated resonances plays a key role in stellar nucleosynthesis. This is particularly true in explosive scenarios such as classical novae and X-ray bursts where resonant capture often dominates the total stellar rate, but level densities are not large enough for statistical capture. In these scenarios, many of the important reactions involve short-lived unstable nuclei, which means that direct measurements of the reaction cross sections in the laboratory require inverse-kinematics techniques. 

For a capture resonance, the two quantities that must be determined are the resonance strength, $\omega\gamma$, and the center-of-mass energy above the relevant particle separation threshold, $E_{cm}$. Together, these determine the isolated contribution of the resonance to the total reaction rate at a given temperature, $T$, according to
\begin{equation}\label{eq:Rate}
	\left<\sigma v\right> \sim \omega\gamma \exp\left[{-E_{cm}/(kT)}\right]/T^{3/2},
\end{equation}
where $k$ is the Boltzmann constant \cite{IliadisBook}. In radioactive beam experiments or stable-beam experiments with low yield, $\omega\gamma$---which is directly proportional to the rate---can typically be determined with total uncertainties on the order of $20$--$30\%.$ The energy appears as an exponential term in Equation~\ref{eq:Rate}, and hence its contribution to the total rate uncertainty depends on the temperature and the absolute value of $E_{cm}$. For classical novae, $T \simeq 0.2$--$0.4$~GK and $E_{cm} \simeq 0.1$--$0.5$~MeV, and uncertainties of a few percent on the resonance energy can translate $20\%$ or more on the reaction rate. 
The impact of resonance energy uncertainties can be even larger in X-ray bursts where temperatures can be as high as $1$~GK and $E_{cm}$ can be $1$ MeV or greater. 
 This exponential magnification of the resonance energy uncertainty underscores the importance of determining $E_{cm}$ for astrophysical capture resonances with total uncertainties of  ${\sim}1\%$ or better.
 
Broadly speaking, resonance energies can be determined in two ways: directly or indirectly. The latter relies on measuring the excitation energy of the resonance together with the relevant particle separation threshold. The best precision typically comes from combining Penning-trap mass measurements to determine the separation threshold  with excitation energies from $\gamma$-ray spectroscopy using high-purity Ge detectors. The combined precision from these measurements can be well below the $1\%$ level in the best cases.
As an example, a resonance energy of $E_{cm} = 0.4808(14)$~MeV ($\Delta E/E = 0.2912\%$) was determined for the key astrophysical resonance in the ${}^{23}\mathrm{Mg}(p,\gamma){}^{24}\mathrm{Al}$ reaction by combining a Penning-trap determination of the \nuc{24}{Al} mass excess, $\Delta = -0.04886(23)$~MeV, together with a $\gamma$-ray spectroscopy determination of the excitation energy, $E_x = 2.3451(14)$~MeV \cite{PhysRevC.92.045803, PhysRevC.77.042802}. 
 For many resonances, determining excitation energies from $\gamma$-ray spectroscopy is not practical, e.g.\ due to low $\gamma$-ray branching ratios. Instead techniques such as missing mass spectroscopy following transfer reactions must be employed, which tend to reduce the precision substantially.

Turning to direct measurements, the traditional way to determine the energies of narrow resonances is to measure the excitation function in a target that is much thicker that the width of the resonance (as well as the energy spread of the beam). With this technique, the energy where the yield falls to $\nicefrac{1}{2}$ of its maximum value is the resonance energy \cite{IliadisBook}. This technique requires yield measurements to be taken at a number of different beam energies, which is usually not practical for inverse-kinematics measurements with radioactive beams, or for stable-beam experiments at large facilities with significant competition for beam time.

An alternative technique has been developed to measure narrow resonance energies in inverse kinematics from the reaction position in an extended gas target. Since the beam continuously loses energy as it passes through the target, the position where the reaction occurs is related to the resonance energy. This method has been successfully used many times at the Detector of Recoils And Gammas Of Nuclear reactions (DRAGON) facility at TRIUMF, with the reaction position determined from $\gamma$-ray hit patterns in an extended array of bismuth germanate (BGO) scintillators. Specifically, the mean $z$ position of BGO hits, with appropriate corrections, is indicative of the mean resonance position. 
Using this technique, the typical energy uncertainties are on the order of $0.5\%$ for sample sizes of ${\sim} 20$ events \cite{Hutcheon2003, Hutcheon2012}. 
However, the BGO detector size ($\phi = 55.8$ mm) and distance from the beam axis 
makes the method less reliable for low sample numbers. The typical standard deviation of BGO $z$ positions is around 60~mm, which requires approximately $10$ or more detected events to obtain $\Delta E/E \sim 1\%$ precision for an $E_{cm} \sim 0.5$~MeV resonance (see Section~\ref{sec:ResPos}).
Obtaining the required number of events is not always possible in low-yield experiments. The hit-pattern method is also sensitive to the angular distribution of the $\gamma$~rays as well as asymmetries in the detection efficiency. 

\begin{figure}
\centering
\centerline{\includegraphics[width=90mm]{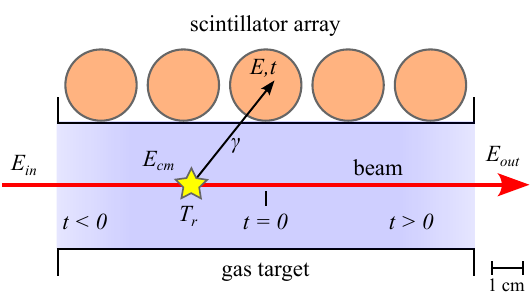}}
\caption{
	Illustration of the resonance timing method. See text for details.
}
\label{fig:ResTime}
\end{figure}

To improve upon the hit-pattern technique, we have developed a new ``resonance timing'' method for determining the position of narrow $(p,\gamma)$ and $(\alpha,\gamma)$ resonances, measured using an extended gas target in inverse kinematics. The method is intended for studying low-lying resonances ($E_{cm} \sim 0.15$--$1.5$~MeV) in medium-mass nuclei ($A\sim 10$--$40$), where the range of inverse-kinematics beam velocities is approximately $v/c = 0.02$--$0.06$. An illustration of the technique is shown in Figure~\ref{fig:ResTime}. The method relies on the incoming beam being delivered in short time bunches, with a measurable time reference signal. We combine the beam reference time with the time of the $\gamma$-ray(s) to determine a time of flight (TOF) relative to the beam bunch crossing the target center. This TOF is directly related to the position where the resonant reaction occurs.
The resonance timing method has the advantage of producing a far narrower spread in inferred position, as compared to the hit-pattern technique.
This allows resonance energies to be determined to the requisite $\lesssim 1\%$ level even with five or fewer detected coincidences. The method is also insensitive to angular distributions and detection efficiency asymmetries. 

For the DRAGON recoil separator, a convenient, non-destructive, beam timing reference is available in the form of the accelerator radio-frequency quadrupole (RFQ) signal. The RFQ has a fundamental frequency of 35.36~MHz and is typically operated at $\nicefrac{1}{3}$ of this value (11.79~MHz, period 84.84~ns) \cite{Fong2001}. Beams are typically delivered to DRAGON with a FWHM time spread of $\leq 4$~ns.  
The resonance timing technique was first demonstrated at DRAGON in an experiment measuring the $E_{cm}=0.475$~MeV resonance in ${}^{22}\mathrm{Ne}(p,\gamma){}^{23}\mathrm{Na}$, using an array of five \labr{} scintillators for $\gamma$-ray detection \cite{HuangThesis}. 
In order to extend the method to future measurements involving radioactive beams, we have installed a mixed array of \labr{}, \cebr{}, and BGO scintillators around DRAGON target location, replacing the traditional array of 30 BGO detectors.  We \edit{performed a first in-beam demonstration of the array,} measuring the $^{23}\mathrm{Na}(p,\gamma){}^{24}\mathrm{Mg}$ reaction. This was part of a broader effort to measure the resonance strength and energy of the ${\sim}0.48$~MeV resonance in the $^{23}\mathrm{Mg}(p,\gamma){}^{24}\mathrm{Al}$ reaction using a \nuc{23}{Mg} radioactive beam. 

The \labr{} and \cebr{} detectors were chosen on the basis of their superior timing properties: ${\sim} 20$~ns decay time and ${\sim} 6$--7~ns rise time, with potential $\gamma$-$\gamma$ timing spread  of ${\sim}0.3$~ns FWHM, as well as excellent energy resolution and good stopping power for ${\sim}$MeV  $\gamma$~rays \cite{Rudiger2020}. The hybrid nature of the array was a consequence of the limited availability of suitable \labr{} or \cebr{} detectors during the \nuc{23}{Mg} beam time. We chose to augment the available \labr{} + \cebr{} detectors with additional BGO detectors in order to maximize detection efficiency. 
 The intent is to use all detected recoil-$\gamma$ coincidences to determine the resonance strength but only $\gamma$~rays detected in the \labr{} or \cebr{} detectors in the resonance timing analysis.


\section{Experiment}
\label{sec:Expt}

For the \edit{first in-beam demonstration utilizing the hybrid scintillator array at DRAGON}, we measured the $^{23}\mathrm{Na}(p,\gamma){}^{24}\mathrm{Mg}$ resonance at $E_{lab}^{(p)} = 0.5121(3)$~MeV ($E_{cm} = 0.4906(3)$ MeV, $E_x = 12.1833(3)$~MeV) \cite{
Endt1973}.\footnote{Later evaluations quote an $E_{lab}^{(p)}$ uncertainty of 0.1~keV \cite{Basunia2022}, but this appears to originate from a misprint in Ref.~\cite{Uhrmacher1985}.
} 
This resonance has a well characterized energy from forward-kinematics measurements, and its energy has also been measured previously at DRAGON using the hit-pattern technique \cite{Hutcheon2012, Erikson2010}. It also occurs in a similar mass and $E_{cm}$ regime as the \nuc{23}{Mg}~+~$p$ resonance that was the focus of the  campaign, making it a good candidate for characterizing the performance of the new technique before \nuc{23}{Mg} experiment. 
Due to different (and heretofore unknown) timing offsets between the detectors, an energy for the $E_{cm} = 0.4906(3)$ MeV \nuc{23}{Na}~+~$p$ resonance could not be extracted from the present data. Instead, the data were used to synchronize the timing offsets between the various detectors, including an overall-array offset that placed the $E_{cm} = 0.4906(3)$ MeV resonance at the expected location in the target. These offsets were then used in the subsequent $^{23}\mathrm{Mg}(p,\gamma){}^{24}\mathrm{Al}$ experiment. 


\begin{figure}
\centering
\centerline{\includegraphics[width=90mm]{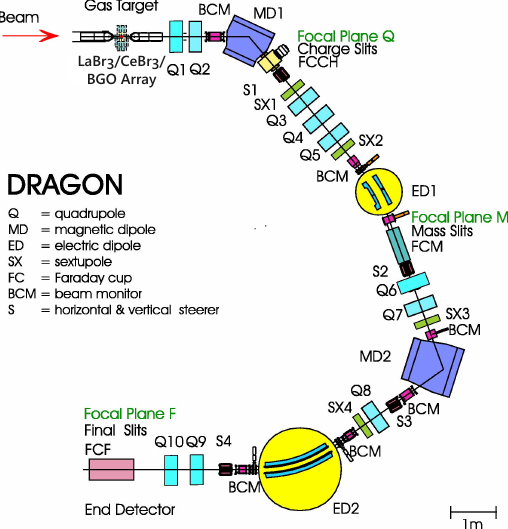}}
\caption{
	\edit{Schematic of the DRAGON recoil separator, including the location of the present \labr{}/\cebr{}/BGO $\gamma$-ray detector array.}
}
\label{fig:Setup}
\end{figure}

The present experiment was scheduled approximately one month before the \nuc{23}{Mg} beam time. The experiment took place in the ISAC-I hall at TRIUMF \cite{Laxdal2003400}, using a modified version of the DRAGON setup \cite{Hutcheon2003}, described below \edit{and shown in Figure~\ref{fig:Setup}}. A beam of stable \nuc{23}{Na}$^{(6+)}$ ions was delivered to DRAGON from the off line ion source \cite{jayamanna:02C711}. The laboratory energy of the beam was measured to be 11.964(17) MeV by centering the beam on slits following DRAGON's first magnetic dipole and calculating the energy using the procedure outlined in Ref.~\cite{Hutcheon2012}, with the recommended magnetic constant value of $c_{mag} = 48.15(7)$~MeV/T$^2$. 
The beam impinged on DRAGON's windowless, recirculating gas target filled with 10.73(6) mbar of H$_2$ gas at a temperature of 306.1(1) Kelvin. The effective length of the gas target is 123~mm. The beam energy was chosen to approximately center the resonance in the target. By comparing the measured beam energies with and without gas in the target, we measured the lab-frame stopping power to be $100.1(3)\times 10^{-15}$~eVcm$^2$, or $5.118(7)$~keV/mm for the present gas pressure and temperature.

The timing signal from the ISAC RFQ (``RF time''), described in Section~\ref{sec:Intro}, was used to define the arrival time of the beam bunches at the DRAGON target.
The beam was delivered in energy-bunched mode, with FWHM energy and time spreads of ${\sim} 0.1\%$ and ${\sim} 4$~ns, respectively. An alternative mode of delivery is time-bunched mode, with FWHM energy and time spreads of ${\sim} 0.4\%$ and ${\sim} 1$~ns.  Although the focus of the experiment was resonance timing, which may suggest time-bunched mode is advantageous, earlier tests with a small \labr{} array showed little difference in overall resonance-timing performance between energy bunched mode vs.\ time-bunched beams. This is due to the broadening of the actual reaction position within the target due to the energy de-focusing in time-bunched mode, which negates the advantages of a tighter time spread.

\begin{figure}
\centering
\centerline{\includegraphics[width=90mm]{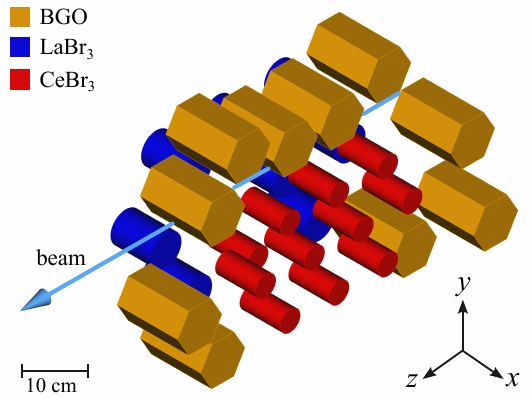}}
\caption{
	Schematic of the $\gamma$-ray detector placement.
}
\label{fig:GammaDetectors}
\end{figure}

The $\gamma$~rays emitted by $^{23}\mathrm{Na}(p,\gamma)^{24}\mathrm{Mg}$ reactions were detected in a scintillator array consisting of 10 BGO detectors from the standard DRAGON setup, 11 cerium bromide (\cebr{}) detectors from the UK National Nuclear Array (NANA) \cite{Shearman2017}, and 9 lanthanum bromide (\labr{}) detectors from the Fast Timing Array (FATIMA) \cite{Rudiger2020}. The BGO detectors had a hexagonal geometry with dimensions of $55.8~\mathrm{mm} \times 76.2~\mathrm{mm}$; the \cebr{} and \labr{} detectors were cylinders with dimensions of $25.4~\mathrm{mm} \times 50.8~\mathrm{mm}$ and $38.1~\mathrm{mm} \times 50.8~\mathrm{mm}$, respectively (all dimensions refer to diameter $\times$ length). Each individual scintillator was coupled to a fast photo-multiplier tube (PMT) for amplification and conversion of the scintillation photons. 
The detectors were energy calibrated using \nuc{60}{Co} and \nuc{244}{Cm}+\nuc{13}{C} sources. The latter produces a 6.13~MeV $\gamma$ ray from the ${}^{13}\mathrm{C}(\alpha,n\gamma){}^{16}\mathrm{O}$ reaction, resulting in both a 6.13~MeV full-energy peak and 5.62~MeV first escape peak for calibration.

The BGO detectors were arranged in a ``crown'' configuration covering the sides and top of the central target position, with the center of each detector (along its long axis) aligned with the beam. Two of the detectors were pulled away from the beam line to accommodate lead shielding put in place for the later \nuc{23}{Mg} experiment. This BGO configuration is a slight modification to the standard placement of the ``crown'' BGO detectors at DRAGON \cite{Hutcheon2003}.
The \labr{} detectors were placed with their long axes perpendicular to the beam axis and their faces flush against the gas target box. They were placed on the right-hand side of the target, when facing in the same direction as the beam. The \cebr{} detectors were placed in a similar configuration on the left-hand side of the target, save for a single detector which was placed on the right side together with the \labr{} detectors (this was later replaced with a \labr{} detector, which arrived at TRIUMF after the present experiment took place). A schematic of the $\gamma$-ray detector setup is shown in \figref{fig:GammaDetectors}, \edit{and a file including the position of each detector (relative to the center of the target) is included as supplementary material.}
%

Recoils from the $^{23}\mathrm{Na}(p,\gamma)^{24}\mathrm{Mg}$ reaction were transmitted through the DRAGON recoil separator, which consists of two stages of magnetic and electric dipoles for $p/q$ and $E/q$ selection, respectively. DRAGON was tuned to accept \nuc{24}{Mg} ions in the $7^+$ charge state. Due to the maximum recoil cone angle of $\leq 17.1$~mrad, ${\sim}100\%$ of \nuc{24}{Mg}$^{(7+)}$ ions were transmitted to the focal plane of DRAGON. The focal plane detectors consisted of a pair of micro channel plates (MCPs), which measured TOF over a local distance of 59~cm, as well as a segmented ionization chamber which measured energy loss of the ions \cite{Vockenhuber2009}. Together with the ``separator'' TOF between the $\gamma$-rays and the recoils detected at the end of DRAGON, the MCP and IC signals provided clear separation and identification of the \nuc{24}{Mg} recoils from background.

While not used in the present analysis, the setup also incorporated a pair of Si surface barrier (SSB) detectors placed at angles of $30^\circ$ and $55^\circ$ relative to the beam, to detect elastically-scattered protons. Together with Faraday-cup readings taken at the beginning and end of each ${\sim} 1$~hour run, these detectors provided an absolute beam intensity normalization. The setup also included a removable plastic scintillator mounted downstream of the gas target. During long experiments, attenuated beam was sent directly into the scintillator at regular intervals to monitor any shifts in the RF time signal. The scintillator data were not used in the present analysis as the data were recorded over a single 1-hour run.

The data acquisition (DAQ) system was the same as in the standard DRAGON setup \cite{Christian2014}. Both the ``head'' ($\gamma$-ray) and ``tail'' (heavy-ion) detectors were fed into free-running DAQ systems with independent triggers recording singles events in either system. Coincidences between $\gamma$~rays and heavy ions were identified offline by comparison of the time-stamps recorded for each head and tail trigger. For the $\gamma$-ray detectors, the anode signals from the individual PMTs were amplified using a fast amplifier, then split into energy and time branches. The energy branches were sent to a 32-channel CAEN V792 charge-to-digital converter (QDC), which integrated the signals within a $750$~ns gate started by a global trigger for the head DAQ.  The timing branches were sent to one of two CAEN V812 16-channel constant fraction discriminators (CFDs). \edit{The CFDs were operated with a delay time of $4$~ns, which is the smallest delay time available for these units without factory modifications.} The OR of all CFD outputs defined the global head trigger, and in addition, the individual channel outputs were sent to a CAEN V1190 time-to-digital converter (TDC), which measured the time of each individual $\gamma$-ray interaction above threshold. Unfortunately, the leading bank of eight TDC signals, 
 all from BGO detectors, were not recorded properly due to a malfunctioning ribbon cable. However, the head trigger signal was also sent to the TDC and was used to recover the timing signals from these eight detectors.

In addition to the standard DAQ, the RF timing signal (described previously) was fed into the TDC of the head system. To avoid swamping the TDC with the rapidly-arriving RF pulses, the RF signal was gated with a copy of the head trigger. The width of the gate was set such that at least three RF pulses entered the TDC for each trigger. As mentioned, the arrival of each RF pulse corresponds to the arrival of a beam bunch at the target, and as such can be taken as a timing reference for determining the beam transit time between the center of the target and the reaction position.

\section{Data Analysis}
\subsection{$^{60}$Co coincidence timing}
\label{subsec:Co60}

Prior to the disassembly of the detector array, an ${\sim}18$ hour run was taken with a ${\sim}38$~kBq \nuc{60}{Co} source. This source emits 1.17 and 1.33 MeV $\gamma$~rays in coincidence, and the run was sufficiently long to observe high-statics $\gamma$-$\gamma$ coincidences in each pair of detectors. This allows us to characterize the array's timing resolution, as well as to synchronize the timing signals from the various detectors. Additionally, we can use the data to estimate the intrinsic energy resolution of the array for moderate $\gamma$-ray energies. Unfortunately, the \nuc{60}{Co} full-energy peaks could not reliably be observed in the BGO detectors during this run. This was due to high thresholds set on these detectors during the \nuc{23}{Mg} beam time which proceeded the source measurement. The thresholds were adjusted online during the \nuc{23}{Mg} beam time until the background count rate was acceptably low, and for the BGO detectors, thresholds ${>}1$ MeV were required.
However, reliable coincidence full-energy peaks were clearly seen for the \labr{} and \cebr{} detectors, and are the focus of the analysis presented in this section.

\begin{figure}
\centering
\centerline{\includegraphics[width=90mm]{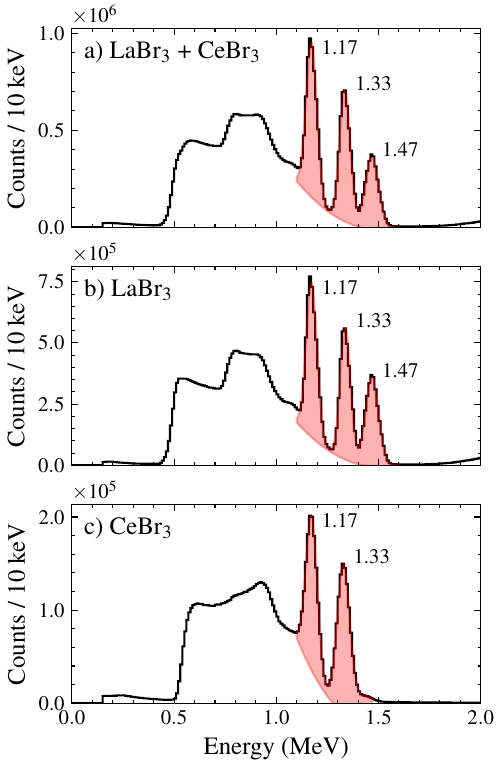}}
\caption{
	Measured \edit{singles $\gamma$-ray} energy spectrum with a \nuc{60}{Co} source for (a) all \labr{} and \cebr{} detectors, (b) all \labr{} detectors, and (c) all \cebr{} detectors. In each panel, the black histograms represent measured data, and the red shaded curves represent the fit results as explained in the text.
}
\label{fig:Co60Energy}
\end{figure}

\begin{figure}
\centering
\centerline{\includegraphics[width=90mm]{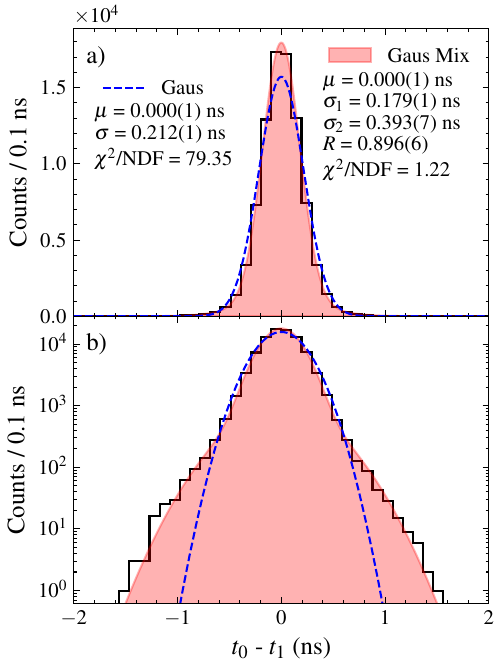}}
\caption{
	\edit{a) Time difference between the 1.33 MeV ($t_0$) and 1.17 MeV ($t_1$) \nuc{60}{Co} full-energy peaks recorded in the \labr{} + \cebr{} detectors (black histogram).  The red shaded and blue dashed curves show the result of Gaussian mixture and Gaussian fits, respectively. The fit parameters for each are indicated in the legends. b) Same as panel (a) but with a logarithmic scale on the $y$ axis.}
}
\label{fig:Co60}
\end{figure}

We first consider the energy spectrum of the fast-timing \labr{} and \cebr{} detectors. \edit{This is shown in Figure~\ref{fig:Co60Energy}, which displays a histograms of the singles $\gamma$-ray energies accumulated across all \labr{}/\cebr{} detectors in the array (panel a), all \labr{} detectors (panel b), and all \cebr{} detectors (panel c).} Three full-energy peaks are clearly evident, namely the 1.17 and 1.33 MeV \nuc{60}{Co} peaks, as well as the 1.47 MeV peak from the \nuc{138}{La} internal radiation \cite{Alzimami2008}. To characterize the energy resolution of the array, we fit the local region of the spectrum near the peaks with the sum of three Gaussian curves on top of a $2^{nd}$ order polynomial background. For the combined \labr{} + \cebr{} array, the respective FWHM energy resolutions at 1.17, 1.33, and 1.47 MeV are 5.42(1)\%, 4.69(1)\%, and 5.68(1)\%. 
For the LaBr$_3$ detectors alone, the resolutions are 5.40(1)\%, 4.48(1)\%, and 5.55(1)\%. For the \cebr{} detectors, the 1.17 and 1.33 MeV resolutions are 5.74(1)\% and 5.51(1)\%, respectively.  In all cases, the energy resolution shows the expected decreasing trend as energy increases from $1.17 \rightarrow 1.33$~MeV. The resolution of the 1.47 MeV peak is broader than the others. This is due to its origin as the sum of a 1.436 MeV $\gamma$~ray with a 0.032 MeV X-ray, leading to a broadening of the total sum peak \cite{Alzimami2008}.

For the \labr{} detectors, the energy resolutions are overall poorer than those previously observed for FATIMA detectors, e.g.\ the ${\sim}2.5\%$ FWHM resolution observed at these energies in Ref.~\cite{Regis2014}. 
This decrease in resolution can  be attributed to the use of charge-integrating ADCs for the energy measurement, \edit{with a common 750~ns gate width for all detectors. The common gate width was a requirement of the CAEN V792 ADC used during the experiment, which only accepts a single gate input for all 32 measurement channels. The 750 ns gate width was chosen to capture the complete decay of the BGO signals; however, it is substantially longer than the length of the \labr{}/\cebr{} detectors, which degrades their energy resolution.  Additionally, the photomultiplier tube high voltages were not optimal for energy resolution:} the PMTs were run at lower voltage than normal to allow removal of 0.44~MeV $\gamma$-ray background, from \nuc{23}{Mg} decay, with discriminator thresholds.

Turning to the $\gamma$-$\gamma$ coincidence timing analysis, we first synchronized the timing offsets between each of the detectors. We did this by choosing detector 20 to be the reference detector, as it was near to the source and had the highest number of \nuc{60}{Co} full-energy peak counts. We then placed a gate on the $\gamma$-$\gamma$ full-energy peak coincidences between each \labr{} or \cebr{} detector and detector 20. For events inside this gate, we determined the mean time difference between the two full-energy peak signals. We then shifted the time signal of each detector by its respective mean value. This allowed us to construct a combined time-difference spectrum for $\gamma$-$\gamma$ full-energy peak coincidences, which was again shifted by the combined mean value such that the combined mean time difference equals zero. The resulting time difference histogram is shown in Figure~\ref{fig:Co60}. An attempt to fit this spectrum with a Gaussian function resulted in a mean and standard deviation of 0.000(1) ns and 0.212(1) ns, respectively. The fit quality is poor, with $\chi^2/\mathrm{NDF} = 79.3$ (here and throughout the paper, $\chi^2$ refers to the Poisson $\chi^2$ as defined in Ref.~\cite{Baker1984}). It is also a poor match visually, as shown by the blue dashed curve in Figure~\ref{fig:Co60}. We instead found that the shape of this spectrum is well described by a Gaussian mixture function,
\begin{equation}
\label{eq:GausMix}
	f(t) = R\mathcal{N}(t;\mu,\sigma_1) + (1-R)\mathcal{N}(t;\mu,\sigma_2),
\end{equation}
where $N(t;\mu,\sigma)$ represents a Gaussian probability density function with mean $\mu$ and standard deviation $\sigma$. $R$ determines the relative strength of each Gaussian and is constrained to be between zero and unity. A binned maximum likelihood fit to the data using Equation~\ref{eq:GausMix} results in $\mu = 0.000(1)$ ns, $\sigma_1 = 0.179(1)$~ns, $\sigma_2 = 0.393(7)$~ns, and $R=0.896(6)$, with $\chi^2/\mathrm{NDF} = 1.22$ ($p = 0.12$). The fit result is shown as the red shaded curve in the figure. 
\edit{The non-Gaussian shape is the result of symmetric extended tails in the data, which are clearly visible in the semilog plot shown in Figure~\ref{fig:Co60}(b). Similar tails are also present in individual detector-to-detector timing spectra. We have not determined a definitive physical cause for these tails; however, we note that they are more prominent in detectors farther away from the source, in particular those on the opposite side of the gas target. This leads us to suspect that the tails originate from scattering of $\gamma$~rays prior to their detection.}
The overall FWHM of \edit{the fitted} Gaussian mixture distribution is 0.443(2)~ns. There is no analytical expression for the FWHM, so we obtained nominal value by numerically by solving for the point where the distribution reaches $\nicefrac{1}{2}$ of its peak value. We estimated the uncertainty using a Monte Carlo technique, varying each of the parameters according to a Gaussian distribution with respective standard deviations equal to the quoted uncertainties. We then took the standard deviation of the resulting FWHMs as the FWHM uncertainty. \edit{The present 0.443(2)~ns FWHM is comparable with, although somewhat larger than, e.g.\ the 0.32 ns FWHM reported for the FATIMA detectors in Ref.~\cite{Rudiger2020}. Possible reasons for the decreased timing resolution include the non-optimal PMT voltages, which degrade timing as well as energy resolution, as well as the $4$~ns CFD delay times, which are slightly longer than the expected rise times of the fast \labr{}/\cebr{} signals. We emphasize that these were necessary operating conditions in order to integrate the array into the existing DRAGON data acquisition system and to operate the array in the later experiment with \nuc{23}{Mg} radioactive beam. We also emphasize that the relatively minor degradation in resolution (as compared to the FATIMA setup) is small compared to the ${\sim}4$~ns time spread of the beam.}

\edit{In addition to the fit to the time-difference spectra across all \labr{}/\cebr{} detectors, we also fit the time-difference spectra of only the \labr{} and \cebr{} detectors. For the \labr{} detectors, the double-Gaussian fit gives $R=0.704(39)$, $\sigma_1=0.204(7)$~ns, $\sigma_2 = 0.420(16)$~ns (FWHM 0.538 ns). For the \cebr{} detectors we obtained $R=0.900(31)$, $\sigma_1=0.175(1)$~ns, $\sigma_2 = 0.350(119)$~ns (FWHM 0.433 ns). These results point to a superior timing resolution for the \cebr{} detectors, which we tentatively attribute to the smaller size of these detectors, which decreases timing spread due to light propagation time.}


\subsection{Recoil Identification}

\begin{figure}
\centering
\centerline{\includegraphics[width=90mm]{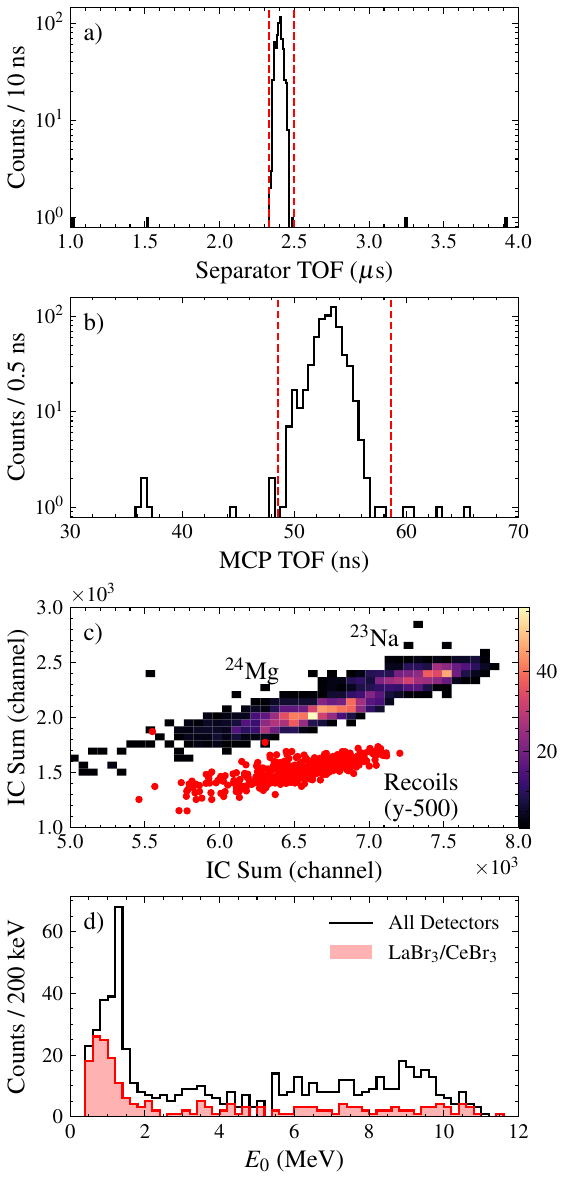}}
\caption{
	Summary of the recoil identification and selection. Panel (a) shows a histogram of separator TOF, with the dashed vertical lines indicating the recoil selection cut made on this parameter. Similarly, panel (b) shows MCP TOF for coincidence events, together with the selection cut. Panel (c) shows IC energy loss in anode 1 vs.\ the sum of all anodes. The color map displays singles events, and the scatter plot displays recoil-$\gamma$~ray coincidences passing the cuts shows in panels (a) and (b). For the scatter plot, 500 channels have been subtracted from the anode 1 signal, so that both the singles and coincidence distributions are clearly visible. \edit{Panel (d) shows histograms of the energies of the highest-energy $\gamma$-ray hit, for all events tagged as recoils; the solid black curve displays the energies for all detectors, and the red shaded curve shows only the \labr{}/\cebr{} detectors.}
}
\label{fig:PID}
\end{figure}

Recoils from the ${}^{23}\mathrm{Na}(p,\gamma){}^{24}\mathrm{Mg}$ reaction were identified and selected based on MCP TOF, separator TOF (effectively the time difference between the $\gamma$-ray and the upstream MCP), and energy loss in the IC anodes. The recoil identification and selection is displayed in Figure~\ref{fig:PID}. Panels (a) and (b) of the figure show histograms of separator and MCP TOF, respectively, each for events tagged as recoil-$\gamma$~ray coincidences based on trigger time-stamp matching. Both distributions displays a clear and prominent peak consisting of recoils, on top of a minimal background from random coincidences between $\gamma$~rays and ``leaky'' $^{23}$Na beam reaching the MCPs. The recoil coincidence events were selected by an AND of cuts placed around the peaks of both the separator and MCP TOF. The actual extent of the cuts is displayed by the dashed vertical lines in panels (a) and (b) of the figure.

While not used in the final analysis, the IC anode signals were useful in confirming the recoil identification. Panel (c) of Figure~\ref{fig:PID} displays the energy loss in the IC ``anode 1'' (second-most upstream) vs.\ the sum of all anodes. Here the color-map represents IC singles events, where two loci resulting from \nuc{24}{Mg} recoils and \nuc{23}{Na} leaky beam are evident. The identification of these two distributions was confirmed by plotting the same distribution for coincidence events passing the separator and MCP TOF cuts (red circles in the figure, which are shifted down by 500 channels on the $y$ axis for display purposes). 
These coincidence recoil events are clearly all clustered in the lower-left locus, confirming that it consists of \nuc{24}{Mg} recoils.

\edit{Figure~\ref{fig:PID}(d) shows the spectrum of highest-detected $\gamma$-ray energies for events tagged as recoils, both for the complete array (solid black histogram) and the \labr{}/\cebr{} detectors alone (shaded red histogram). The shape of the energy spectrum is commensurate with the $\gamma$-ray decay scheme for the $E_{lab}^{(p)} = 0.5121(3)$~MeV resonance, which has a \nuc{24}{Mg} excitation energy of 12.1833~MeV and decays primarily through a 10.812~MeV $\gamma$~ray to the $2^+$, 1.369~MeV first excited state \cite{Basunia2022}.}
 From the ratio of coincidence to singles recoil events, we are able to estimate the efficiency of the $\gamma$-ray detector array. To do this, we selected recoil events in singles by combining a gate on the \nuc{24}{Mg} region of the IC, shown in Figure~\ref{fig:PID}(c), together with the MCP TOF gate shown in Figure~\ref{fig:PID}(b). Taking the ratio of coincidence to singles recoils results in a total detection efficiency of $45.4(2.2)\%$ for the complete array, $16.3(1.3)\%$ for the \labr{}/\cebr{} detectors, and $29.1(1.7)\%$ for the BGO detectors. These efficiencies represent the probability of detecting any $\gamma$~ray from the ${}^{23}\mathrm{Na}(p,\gamma){}^{24}\mathrm{Mg}$ reaction with a deposited energy above the detector threshold; \edit{the relevant energy distribution for the efficiency is that shown in Figure~\ref{fig:PID}(d).}

\subsection{RF Timing Analysis}
\label{sec:RfTime}

\begin{figure}
\centering
\centerline{\includegraphics[width=90mm]{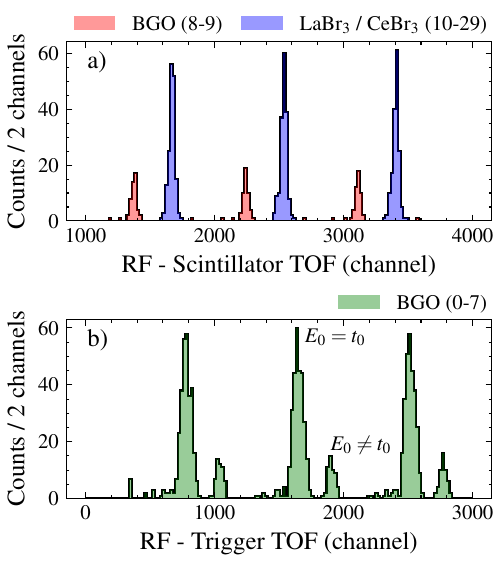}}
\caption{
	Raw RF timing signals. Panel (a) shows a histogram of the difference in TDC channel between the RF signal and the individual timing signal of the $E_0$ (largest-energy) $\gamma$-ray detector. As indicated in the legend, events in the histogram are sub-divided into those where the $E_0$ detector is a \labr{} or \cebr{} (blue) or BGO (red). As explained in the text, BGO detectors 0--7 are excluded from this plot. Panel (b) shows the difference in TDC channel between the RF signal and the head trigger signal, for BGO detectors 0--7.
}
\label{fig:RFraw}
\end{figure}

In order to characterize the RF timing properties of the scintillator array, we analyzed the time signals from recoil events in detail. The product of this analysis is an RF--$\gamma$-ray TOF spectrum from which we can extract a mean value and associated confidence interval. This mean TOF can then be translated into reaction position within the extended gas target, and from there into resonance energy.

Histograms of the raw RF--$\gamma$-ray TOF are shown in Figure~\ref{fig:RFraw}. Here panel (a) displays the difference in TDC channel between the RF pulse and the CFD pulse from the $\gamma$-ray detector with the largest deposited energy (the ``$E_0$'' $\gamma$-ray). Recall that for BGO detectors 0--7, the TDC signals were not operating properly; hence, panel (a) excludes events where the $E_0$ detector is one of these. The periodic structure of the RF signal is clearly evident in the histogram, with three different TOF peaks present with a separation corresponding to the $84.84$~ns RF period. The difference in peak location between the \labr{}/\cebr{} (blue-shaded) and BGO (red-shaded) detectors is due to differing signal processing and transit times between the two types of detector; these time differences were subsequently corrected for as described later in this section.

Panel (b) of Figure~\ref{fig:RFraw} shows the difference in TDC channel between the RF signal and the global head trigger, for events where the $E_0$ detector is one of BGO detectors 0--7. The head trigger is effectively a copy of the OR of all CFD channels, and  it is not impacted by the TDC problem for these detectors. In this spectrum, the periodic nature of the RF timing is again evident; however, for each RF bunch there are two peaks. The main (higher-intensity) peak consists of ``good'' events where the $E_0$ detector is also the detector that defines the trigger (hence $E_0 = t_0$ as labeled in the figure). The lower-intensity peak is due to events where the $E_0$ detector is not the trigger ($E_0\neq t_0$)---in particular, events where a \labr{} or \cebr{} detector is in coincidence with one of the BGOs, with the BGO recording more deposited energy. When this happens, the \labr{}/\cebr{} signal will be generated first and hence define the global trigger. For subsequent analysis, we removed these $E_0\neq t_0$ events and considered only those where the $E_0$ detector was also the detector defining the trigger.

Due to differences in signal processing and transit time, e.g., due to different detector properties or cable lengths, the individual RF signals must be synchronized to construct a timing spectrum for the complete array. In order to synchronize the various detector signals, we determined the mean RF TOF (of the middle peak in Figure~\ref{fig:RFraw}) for each individual detector, then shifted that detector's time signal by the mean value to construct a combined time spectrum with a mean at channel zero. The choice to place the peak at zero is arbitrary. For a typical experiment measuring unknown resonance energies, the TOF centroids would first be determined for a known resonance like the one presently under study. The synchronized spectrum would then be shifted to the TOF value corresponding to the known resonance position within the target. These timing shifts would be preserved for subsequent measurements with unknown resonance energies.

After synchronizing the RF timing spectra such that they all have a mean of 0 channels, the separation between the three RF TOF peaks was used to calibrate channel number into ns. The CAEN V1190 TDCs used during this experiment have a nominal slope of approximately 0.1~ns/channel. 
For the calibration, we found an average difference of $868.3$ channels between the mean value of each of the synchronized RF TOF peaks. Dividing into the RF period of $84.84$~ns results in a slope of $0.09771$~ns/channel. For subsequent analysis, the synchronized RF TOF signals were multiplied by this slope in order to give a TOF in nanoseconds.

\begin{figure}
\centering
\centerline{\includegraphics[width=90mm]{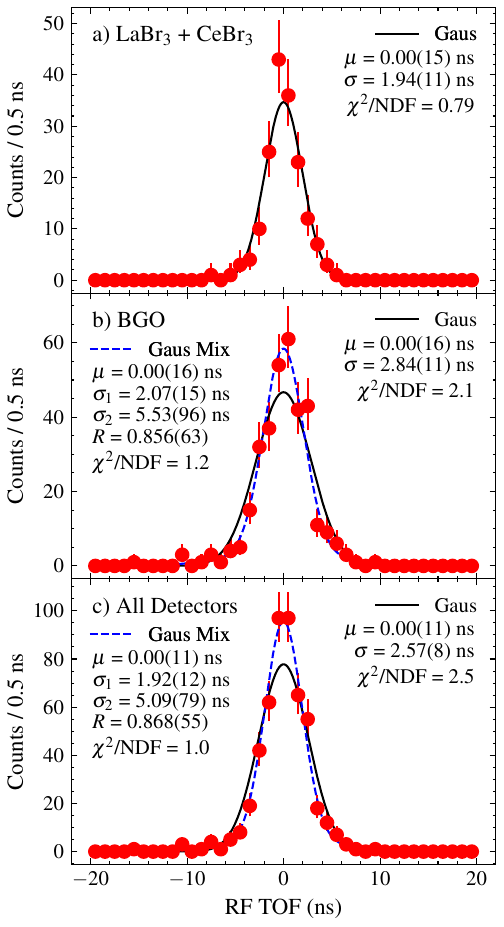}}
\caption{
	Histograms of the RF TOF (filled red circles with error bars), along with Gaussian fits (black solid curves) and Gaussian mixture fits (blue dashed curved). Best-fit parameters are indicated in the figure legends. The various panels show data and fits from different sub-sets of the detector array: a) all \labr{} and \cebr{} detectors, b) all BGO detectors, and c) the complete array.
}
\label{fig:RFTOF_all3}
\end{figure}

After the RF timing synchronization and calibration, we analyzed the RF TOF distributions, looking separately at the fast-timing \labr{}/\cebr{} detectors, the BGO detectors, and the complete array. For the \labr{}/\cebr{} detectors, a histogram of the RF TOF is shown  in Figure~\ref{fig:RFTOF_all3}(a). A maximum likelihood fit of the data with a Gaussian distribution gives good agreement, with $\mu = 0.00(15)$~ns, $\sigma = 1.94(11)$~ns (FWHM 4.57(26) ns), and $\chi^2/\mathrm{NDF} = 0.79$ ($p = 0.72$).
%

As expected, the BGO detectors, shown in Figure~\ref{fig:RFTOF_all3}(b), have a broader time distribution than the \labr{}/\cebr{}. However, the difference is not substantial due to the dominance of the time spread of the resonance itself over the detector timing resolution. For the BGO detectors, a Gaussian fit gives $\mu = 0.00(16)$~ns and $\sigma = 2.84(11)$~ns but is a poor match to the data with $\chi^2/\mathrm{NDF} = 2.1$ ($p=4.4\times10^{-3}$). A Gaussian mixture (Equation~\ref{eq:GausMix}) with $\mu = 0.00(16)$~ns,  $\sigma_1 = 2.07(15)$~ns, $\sigma_2 = 5.53(96)$~ns, and $R=0.856(63)$ proves a good match with $\chi^2/\mathrm{NDF} = 1.2$ ($p=0.30$). This distribution has a total FWHM spread of 5.05(38)~ns, obtained numerically using the same procedure described in  Section~\ref{subsec:Co60}.

Finally, the complete detector array, Figure~\ref{fig:RFTOF_all3}(c), has a width in between that of the \labr{}/\cebr{} and BGO detectors. A Gaussian fit gives $\mu = 0.00(11)$~ns and $\sigma = 2.57(8)$~ns but is a poor match with $\chi^2/\mathrm{NDF} = 2.5$ ($p = 4.1\times 10^{-4}$). Again the Gaussian mixture, with $\mu = 0.00(11)$~ns,  $\sigma_1 = 1.92(12)$~ns, $\sigma_2 = 5.09(79)$~ns, and $R=0.868(55)$ (FWHM 4.67 ns), is a good fit to the data with $\chi^2/\mathrm{NDF} = 1.0$ ($p=0.44$). This Gaussian mixture has a total FWHM of 4.67(31)~ns.

\section{Expected Performance in Low Statistics Experiments}
\label{sec:Bootstrap}

The primary application of the resonance timing method is to radiative capture experiments with modest-intensity radioactive beams (down to ${\sim}10^{6}/$s), or in stable-beam measurements with low cross sections. Typically these experiments involve very low counting rates, with only a handful of events collected over the course of an experiment. A major advantage of the resonance timing method is its ability to give robust  resonance energy measurements even when the total number of coincidence counts is well into the single digits. This is due to the relatively tight spread in the RF TOF distribution which results in a relatively small uncertainty on the mean time even with low statistics.  This is a major advantage of the technique over the hit pattern analysis typically employed at DRAGON, which requires $\gtrsim{10}$ events for a reliable analysis.

In order to estimate the performance of the resonance timing method for a realistic low-count experiment, we have used a bootstrap technique to perform a series of pseudo-experiments that estimate the distribution of mean RF times and confidence intervals for differing numbers of events. The pseudo-experiments were performed as follows: for a given number of total counts $N$, we drew 500,000 random samples, each of size $N$, from the RF TOF data presented in Section~\ref{sec:RfTime}. The samples were drawn with replacement, meaning a given event can be selected more than once within an iteration. For each random sample, we determined the resonance time, $T_r$, as being equal to the mean RF TOF. We also determined the $1\sigma$ confidence interval on the resonance time, $\sigma_{T_r}$, using two different techniques, explained below. The resulting pseudo-data thus give an estimate of the magnitude of $\sigma_{T_r}$ to be expected from an experiment, as well as the reliability of the method used to estimate $\sigma_{T_r}$. We performed separate  sets of pseudo-experiments using the \labr{}/\cebr{} data, the BGO data, and data from all detectors, for sample sizes $N = 3,$ 5, 10, 25, 50, and 100. We also repeated the pseudo-experiments using only data where the $E_0$ $\gamma$-ray energy was below $2$~MeV. This simulates an experiment where only low-energy $\gamma$~rays are detected, e.g.\ due to a small $Q$ value for the radiative capture reaction under study.

\begin{figure*}
\centering
\centerline{\includegraphics[width=190mm]{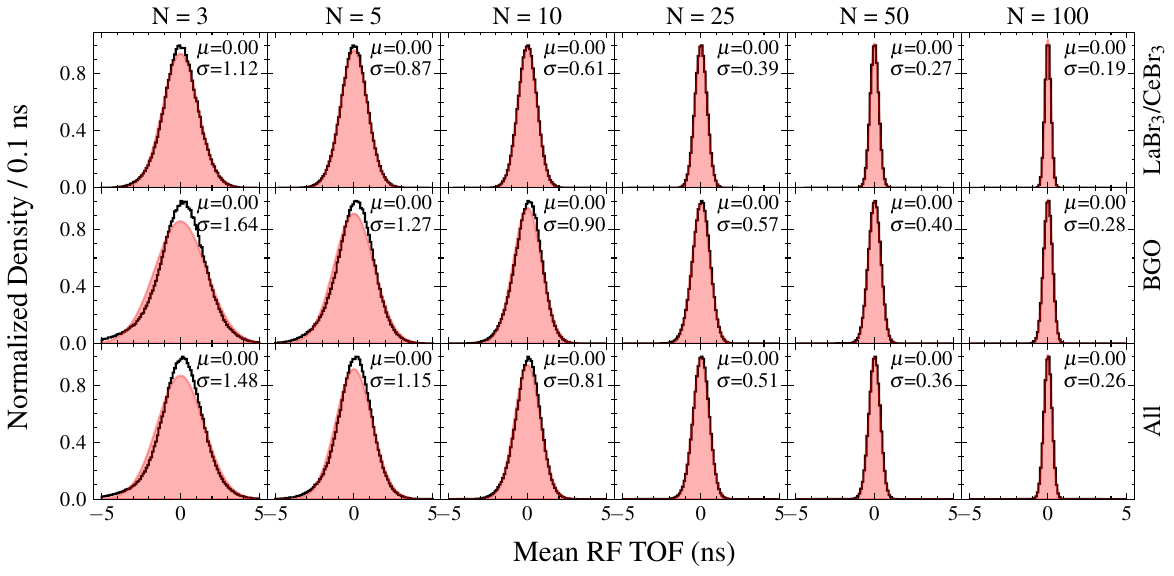}}
\caption{
	Distribution of resonance times obtained from the bootstrap pesudo-experiments performed using all $\gamma$-ray energies. In each panel, the black solid curves show a histogram of the mean RF TOF values calculated across 500,000 pseudo-experiments (for display purposes, all histograms are normalized such that the maximum bin content is equal to unity). The shaded red curves show the results of a Gaussian estimation, with mean and standard deviation indicated in the figure legends. Each column shows the results for different sample sizes, as labeled on the top of the figure, while each row shows results from a different sub-set of detectors (labeled on the right hand side of the figure).
}
\label{fig:bootstrap_results}
\end{figure*}

\begin{figure*}
\centering
\centerline{\includegraphics[width=190mm]{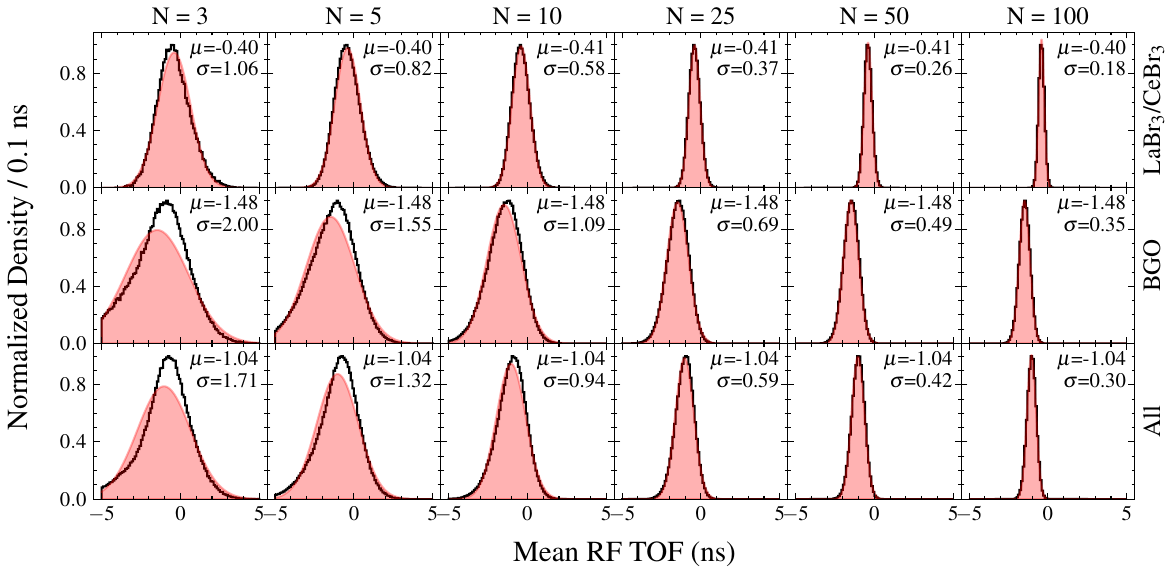}}
\caption{
	Same as Figure~\ref{fig:bootstrap_results} but only including events where the highest-energy $\gamma$-ray hit is below 2~MeV.
}
\label{fig:bootstrap_results_2}
\end{figure*}

\begin{figure*}
\centering
\centerline{\includegraphics[width=190mm]{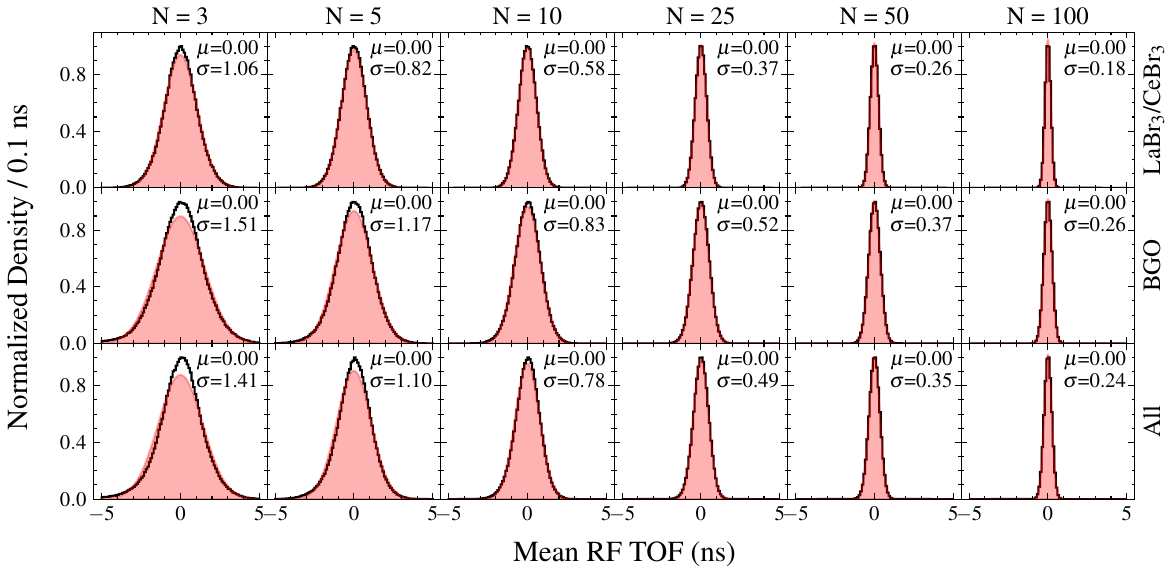}}
\caption{
	Same as Figure~\ref{fig:bootstrap_results} but including a walk correction (see text for details).
}
\label{fig:bootstrap_results_3}
\end{figure*}

\edit{Histograms of the} resulting distributions of resonance times from the various pseudo-experiments are shown \edit{as the solid black curves} in Figure~\ref{fig:bootstrap_results} (all energies), Figure~\ref{fig:bootstrap_results_2} ($E_0<2$~MeV), \edit{and Figure~\ref{fig:bootstrap_results_3} (all energies, walk corrected; see below)}. 
\edit{In addition to the histograms, we also show their Gaussian estimations as the shaded red curved in the figures.}
For the trials performed with all energies, the time distributions are reliably centered at zero (as expected, since the original distribution from which the samples are taken has a mean of zero). There is a small deviation from the Gaussian shape at low sample numbers, especially for the BGO detectors, but the distributions converge to Gaussian by $N=10$, as expected from the central limit theorem.
For the $E_0<2$~MeV trials, there is a clear bias towards negative times, with the sample mean around $-0.4$~ns for the \labr{}/\cebr{} detectors, $-0.55$~ns for the BGOs, and $-0.49$~ns for all detectors. The small-sample number distributions are also significantly non-Gaussian for the BGO detectors.

The shift towards negative times when $E_0<2$~MeV is not surprising when considering that the mean RF TOF values for the original data (with $E_0<2$~MeV) are $-0.405$~ns, $-1.48$~ns, and $-1.04$~ns for the \labr{}/\cebr{}, BGO, and all detectors respectively. This is likely due to the presence of a modest timing walk, which introduces correlations between $E_0$ and $T_r$. 
This shift of the mean TOF for different $E_0$ energies highlights the importance of carefully considering energy-time correlations when calibrating detectors for an experiment. For example, if the presently-used RF timing offsets were used in an experiment where the detected $\gamma$-rays are all below 2~MeV,  the resulting mean recoil time would show a bias of around $-0.5$~ns. 

\edit{
	Since both the time and energy data are available on an event-by-event basis, it is possible to perform a walk correction to counter this effect. We have done this with the present data by dividing the $E_0$ data into 1~MeV bins and shifting the times in coincidence with each energy bin by their corresponding mean value (for example, the mean uncorrected time for events with $0 \leq E_0 < 1$~MeV is $-1.297$~ns, so the walk correction adds $1.297$~ns to all recorded times with energies in this range). We performed separate walk corrections for the \labr{}/\cebr{} detectors, the BGOs, and all detectors. The standard deviation of the various energy-bin shifts was 0.65~ns for the \labr{}/\cebr{} detectors, 1.4~ns for the BGO detectors, and 0.70~ns for the complete array; these numbers give an indication of the systematic uncertainty that would be introduced if energy-time correlations were ignored completely.  Using the walk-corrected data we ran another set of bootstrap pseudo-experiments, whose results are shown in Figure~\ref{fig:bootstrap_results_3}. As expected, the mean walk corrected times are all equal to zero. The standard deviations are smaller than the corresponding trials without walk correction---especially for the BGO detectors. This is the result of timing spread that is introduced into the non-corrected data by the walk effect; the walk correction effectively eliminates this spread and results in a smaller standard deviation of times across the various trials.
}


In addition to inspecting histograms of the resonance times, we also estimated the $1\sigma$ confidence intervals on the resonance times obtained in each of the pseudo experiments, $\sigma_{T_r}$. These correspond to the uncertainty obtained on the resonance time value, i.e.\ the typical report of an experiment would be $T_r \pm \sigma_{T_r}$.
For this analysis, we calculated the confidence intervals using two different methods. For each method, we report the mean and standard deviation of $\sigma_{T_r}$ in Table~\ref{tab:Bootstrap}. We also report the coverage probability, $P_c$, of the various methods---that is, the fraction of confidence intervals in each pseudo-experiment that enclose the true mean value. For $1\sigma$ intervals, we expect $P_c = 68.3\%$. If the coverage is smaller than this, it indicates that $\sigma_{T_r}$ is underestimated, while if the coverage is greater it indicates $\sigma_{T_r}$ is overestimated. 

The common technique of binning the data and performing a maximum likelihood or least-squares fit to extract the mean time and its associated confidence interval is not well suited to the very small sample sizes considered here, due to the strong impact of bin size on the results. This would also be true for a real experiment detecting a small number of events. Instead, we rely on estimation techniques based on the original, unbinned data, which also have the advantage that they make no assumptions about the underlying distribution of the data.
The first technique is to use the standard estimator of the error on the mean, $\sigma_{T_r} = s_{T_r}/\sqrt{N}$. Here, $N$ is the number of samples and $s_{T_r}$ is the sample standard deviation given by
\begin{equation}
\label{eq:stddev}
	s_{T_r} = \sqrt{
		\frac{1}{N-1} \sum_{i=1}^{N} (T_i - T_r)^2,
	}
\end{equation}
with $T_r$ the mean TOF values across all $N$ samples \cite{Lyons1986}.
We label this the ``Gaus'' method in the tables because it corresponds to the limiting case where the deviation between the sample mean of the data and the true mean follows a Gaussian distribution with mean 0 and standard deviation $s_{T_r}/\sqrt{N}$.
This method is correct when the $s_{T_r}$ is close to the true standard deviation, which occurs for large $N$. However, for small $N$,  $s_{T_r}$ can be significantly smaller than the true value. This is borne out in the results of the pseudo-experiments, where the coverage for small sample sizes is significantly below $68.3\%$. This indicates that this technique is a poor choice for estimating $\sigma_{T_r}$ in an experiment with a small number of detected events and is likely to under-estimate the uncertainty on the resonance time (and by extension the resonance position and energy).

A better method for estimating $\sigma_{T_r}$ is to recognize that the difference between the mean estimated from the data and the true mean, for a sample size $N$, is distributed according to $\mathcal{T}_{N-1} s_{T_r}/\sqrt{N}$, where $\mathcal{T}_{N-1}$ represents a Student's T distribution with $N-1$ degrees of freedom \cite{James2006}.  As a result, the $1\sigma$ confidence interval on the mean TOF is given by $\sigma_{T_r} = As_{T_r}/\sqrt{N}$, where $2\int_{0}^{A} \mathcal{T}_{N-1}(x) \mathrm{d}x$ = 0.683. As shown in Table~\ref{tab:Bootstrap}, in the columns labeled ``T'', there is only a very modest under-coverage of $<3\%$ for the smallest sample sizes. Using this method, the confidence intervals on $T_r$ are larger than the Gaussian approximation, starting at an average of $1.28$~ns for three events (with a standard deviation of $0.75$~ns), and converging to the Gaussian approximation for sample sizes of around $N=25$. The good coverage properties of the ``T'' method of confidence interval estimation make it the preferred choice for analyzing experiments with $\lesssim 20$ detected events.

\edit{
The results of the pseudo-experiments indicate that the \labr{}/\cebr{} detectors out-perform the BGO detectors, with the walk-corrected $\sigma_{T_r}$ (using the ``T'' method) being around $40\%$ larger for the BGO detectors vs.\ the \labr{}/\cebr{} detectors at the same sample number. While significant, this difference is modest compared with the factor ${\sim}14$ larger intrinsic timing resolution of the BGO detectors\footnote{\edit{
	We estimated a FWHM resolution of ${\sim}4.9$~ns for the BGO detectors by combining the single-Gaussian fit results presented in Figure~\ref{fig:RFTOF_all3} with the ${}^{60}\mathrm{Co}$ coincidence timing results presented in Figure~\ref{fig:Co60}. The estimated BGO single-detector FWHM resolution is given by $\mathrm{FWHM}_{BGO} = 2.355 {\sqrt{\sigma_2^2 - (\sigma_1^2 - \sigma_{d1}^2)}}$, with $\sigma_1(\sigma_2) = 1.94(2.84)$~ns the \labr{}/\cebr{} (BGO) RF TOF resolutions from Figure~\ref{fig:RFTOF_all3} and $\sigma_{d1} = 0.212/\sqrt{2}$ the single-detector \labr{}/\cebr{} timing resolution from Figure~\ref{fig:Co60}.
}}, which highlights the dominance of the beam timing spread over the detector resolution on the overall resonance timing uncertainties. 
When designing an experiment, the advantages of improved detector timing resolution must be weighed against potential disadvantages in detection efficiency. For example, if using BGO detectors results in an expected $5$ resonance counts, vs.\ $3$ for \labr{}/\cebr{} detectors, the advantages of the faster timing would essentially be eliminated by the better statistics provided by the BGO detectors.
}

\begin{table*}
\begin{center}
\caption{Summary of the results of the bootstrap pseudo-experiments described in the text. In the columns labeled $\sigma_{T_r}$, the values outside the parenthesis represent the mean confidence interval on the resonance time across all bootstrap iterations. The values inside the parenthesis represent the standard deviation on the same. Columns labeled $P_c$ show the coverage probability of the corresponding confidence intervals. Results are presented using both the ``Gaus'' and ``T'' methods of obtaining confidence intervals (described in the text), and for various sub-sets of the data, noted in the column labels.
}
\label{tab:Bootstrap}
\centerline{
\begin{tabular}{c|cc|cc|cc|cc|cc|cc}
\cline{2-13}
\multicolumn{1}{c}{} & 
\multicolumn{4}{c|}{All Energies} & 
\multicolumn{4}{c|}{$E_0 < 2$ MeV} & 
\multicolumn{4}{c}{Walk Corrected} \\
\cline{2-13}
\multicolumn{1}{c}{} & 
\multicolumn{2}{c|}{Gaus} & 
\multicolumn{2}{c|}{T} & 
\multicolumn{2}{c|}{Gaus} & 
\multicolumn{2}{c|}{T} &
\multicolumn{2}{c|}{Gaus} & 
\multicolumn{2}{c}{T} \\
\hline
$N$ & 
$\sigma_{T_r}$ (ns) & $P_c$ & 
$\sigma_{T_r}$ (ns) & $P_c$ & 
$\sigma_{T_r}$ (ns) & $P_c$ & 
$\sigma_{T_r}$ (ns) & $P_c$ & 
$\sigma_{T_r}$ (ns) & $P_c$ & 
$\sigma_{T_r}$ (ns) & $P_c$\\
\hline
\multicolumn{13}{c}{\labr{}/\cebr{}}\\
\hline
  3 & 0.97(56) & 57.0\% & 1.3(7) & 68.6\% & 0.92(53) & 56.4\% & 1.2(7) & 68.1\% & 0.92(54) & 56.4\% & 1.2(7) & 68.1\% \\
  5 & 0.80(34) & 61.4\% & 0.91(38) & 67.4\% & 0.76(31) & 61.1\% & 0.87(35) & 67.1\% & 0.76(32) & 61.1\% & 0.87(37) & 67.1\% \\
 10 & 0.59(17) & 65.0\% & 0.62(18) & 67.7\% & 0.56(15) & 64.9\% & 0.59(16) & 67.4\% & 0.56(16) & 64.9\% & 0.59(17) & 67.4\% \\
 25 & 0.38(7) & 66.9\% & 0.39(7) & 67.9\% & 0.36(6) & 67.1\% & 0.37(6) & 68.1\% & 0.36(6) & 67.1\% & 0.37(7) & 68.1\% \\
 50 & 0.27(3) & 67.6\% & 0.27(3) & 68.1\% & 0.26(3) & 67.7\% & 0.26(3) & 68.2\% & 0.26(3) & 67.7\% & 0.26(3) & 68.2\% \\
100 & 0.19(2) & 67.9\% & 0.19(2) & 68.1\% & 0.18(1) & 67.9\% & 0.18(1) & 68.2\% & 0.18(2) & 67.9\% & 0.18(2) & 68.2\% \\
\hline
\multicolumn{13}{c}{BGO}\\
\hline
  3 & 1.4(9) & 56.6\% & 1.8(1.2) & 68.3\% & 1.7(1.1) & 54.5\% & 2.2(1.5) & 66.6\% & 1.3(8) & 54.5\% & 1.7(1.1) & 66.6\% \\
  5 & 1.1(6) & 61.2\% & 1.3(7) & 67.4\% & 1.4(7) & 59.1\% & 1.6(8) & 65.2\% & 1.1(5) & 59.1\% & 1.2(6) & 65.2\% \\
 10 & 0.84(31) & 64.2\% & 0.89(33) & 66.9\% & 1.0(4) & 63.4\% & 1.1(4) & 66.0\% & 0.79(26) & 63.4\% & 0.83(28) & 66.0\% \\
 25 & 0.55(13) & 66.2\% & 0.56(13) & 67.3\% & 0.68(15) & 66.3\% & 0.69(15) & 67.3\% & 0.51(11) & 66.3\% & 0.52(11) & 67.3\% \\
 50 & 0.40(7) & 67.1\% & 0.40(7) & 67.6\% & 0.48(7) & 67.4\% & 0.49(7) & 67.9\% & 0.37(6) & 67.4\% & 0.37(6) & 67.9\% \\
100 & 0.28(3) & 67.8\% & 0.28(3) & 68.0\% & 0.34(4) & 67.8\% & 0.35(4) & 68.0\% & 0.26(3) & 67.8\% & 0.26(3) & 68.0\% \\
\hline
\multicolumn{13}{c}{All Detectors}\\
\hline
  3 & 1.2(8) & 56.7\% & 1.6(1.1) & 68.7\% & 1.4(1.0) & 56.0\% & 1.8(1.3) & 68.9\% & 1.2(8) & 56.0\% & 1.6(1.0) & 68.9\% \\
  5 & 1.0(5) & 61.4\% & 1.2(6) & 67.6\% & 1.2(6) & 59.5\% & 1.3(7) & 66.0\% & 0.98(49) & 59.5\% & 1.1(6) & 66.0\% \\
 10 & 0.76(28) & 64.3\% & 0.81(30) & 67.0\% & 0.87(34) & 62.8\% & 0.92(36) & 65.5\% & 0.73(26) & 62.8\% & 0.77(27) & 65.5\% \\
 25 & 0.50(12) & 66.3\% & 0.51(12) & 67.3\% & 0.57(14) & 65.8\% & 0.59(14) & 66.9\% & 0.48(11) & 65.8\% & 0.49(11) & 66.9\% \\
 50 & 0.36(6) & 67.2\% & 0.36(6) & 67.7\% & 0.41(7) & 67.0\% & 0.42(7) & 67.5\% & 0.34(6) & 67.0\% & 0.35(6) & 67.5\% \\
100 & 0.26(3) & 67.7\% & 0.26(3) & 68.0\% & 0.29(4) & 67.7\% & 0.29(4) & 67.9\% & 0.24(3) & 67.7\% & 0.24(3) & 67.9\% \\
\hline
\end{tabular}
}
\end{center}
\end{table*}

\section{Resonance Position and Energy}
\label{sec:ResPos}

Once a mean RF TOF and its associated uncertainty have been determined, this needs to be translated into a confidence interval on the resonance energy. In this section, we present equations for determining the resonance position as a function of time as well as the resonance energy as a function of position. These equations account for special relativity and the slowing down of the beam as it passes through the target, but they ignore any changes in the stopping power as the beam traverses the target. They also treat the target as being a volume of constant gas density with effective length $L=123$~mm, the established effective length of the dragon gas target \cite{Hutcheon2003}. This approximation is valid as long as the resonant reaction occurs in the central region of the target. 
Using the equations presented in this section, we can translate the results of Section~\ref{sec:Bootstrap} into expected uncertainties on resonance position and energy for various sample sizes and sub-sets of detectors.

We start  with the equation for relativistic velocity,
\begin{equation}
v/c = \sqrt{1 - \frac{1}{(1+E/M_b)^2}},\\
\end{equation}
where $E$ is the beam kinetic energy (lab frame, MeV), and $M_b$ is the beam mass times $c^2$, in MeV. We can substitute $v=dz/dt$, where $z$ is position along the length of the target (in mm), as well as $E = E_0 - \epsilon z$, where $E_0$ is the incoming beam energy and $\epsilon$ is the stopping power in MeV/mm. This gives an integral equation to obtain time as a function of position:
\begin{equation}
	\int_0^t {dt^\prime}  = 
		\frac{1}{c} \int_0^z \left[
			1 - \frac{1}{[1+(E_0-\epsilon z^\prime)/M_b]^2}
		\right]^{-1/2} dz^\prime.
\end{equation}
The integral can be evaluated analytically and simplifies to 
\begin{equation}
\label{eq:tofz}
t(z) = 
		\frac{
        \sqrt{E_0(2 M_b+E_0)} -
        \sqrt{(E_0-\epsilon z)(2 M_b+E_0-\epsilon z)}
        }{\epsilon c}.
\end{equation}
This can be inverted to give position as a function of time:
\begin{equation}
\label{eq:zoft}
    z(t) = -\frac{1}{2\epsilon}\left(B + \sqrt{B^2 - 4C}\right),
\end{equation}
with
\begin{align*}
B&=-2(M_b+E_0)\\
C&=E_0^2 + 2M_b E_0 - \left(\sqrt{E_0(2M_b+E_0)}-\epsilon c t\right)^2.
\end{align*}
Equations~\ref{eq:tofz} and \ref{eq:zoft} take the zero point of position and time to be the moment the beam crosses the target entrance. A more convenient zero point is the center of the target; this reference shift can be made simply by substituting $E_0\rightarrow (E_0-\epsilon L/2)$. Plugging the resonance time, $T_r$, into Equation~\ref{eq:zoft} gives the resonance position, $Z_r$ (relative to the target center). Once this is obtained, we can calculate the resonance energy as: 
\begin{align}
	E_{lab} &= E_0 - \epsilon(Z_{r} + L/2) \label{eq:Elab}\\	
	E_{cm} &= \sqrt{M_t^2 + M_b^2 + 2M_t(M_b + E_{lab})} - M_t - M_b, \label{eq:Ecm}
\end{align}
with $M_t$ the target mass times $c^2$, in MeV.

From Equation~\ref{eq:zoft}, we can calculate the uncertainty on the resonance position, $\sigma_{Z_r}$ 
\edit{using standard Gaussian uncertainty propagation rules, which in the absence of correlations results in  
\begin{equation}
\sigma_{Z_r} = \sqrt{
	\sigma_{T_r}^2\left(\frac{\partial z}{\partial t}\right)^2
	+ \sigma_{E_0}^2\left(\frac{\partial z}{\partial E_0}\right)^2
	+ \sigma_{T_r}^2\left(\frac{\partial z}{\partial \epsilon}\right)^2
}.
\end{equation}
The terms involving $\sigma_{E_0}$ and $\sigma_\epsilon$ amount to corrections of approximately $4 \times 10^{-6}$ relative to the $\sigma_{T_r}$ term. Neglecting these terms results in a convenient closed-form expression for the uncertainty on the resonance position,
\begin{equation}
\sigma_{Z_r} = \frac{c\left[\sqrt{E_0(E_0+2M_b)} - \epsilon c T_r\right]}{
        \sqrt{(M_b+E_0)^2  - C}
    } \sigma_{T_r}.
\end{equation}
}
%
 A full treatment of uncertainties on the resonance energy requires consideration of correlations between the beam energy, stopping power, and resonance position. Hence, for calculating uncertainties on $E_{cm}$, we use the python \emph{uncertainties} package for the error propagation \cite{Uncertainties}. This package fully treats the correlations between the various parameters using their analytic covariance and correlation matrices.

Using Equations~\ref{eq:zoft}, \ref{eq:Elab}, and \ref{eq:Ecm}, we can calculate the confidence intervals for resonance position and energy from the bootstrap pseudo-experiment results presented in Section~\ref{sec:Bootstrap}. For each bootstrap iteration, we had previously found a confidence interval on the resonance time, which we now convert into a confidence interval on the resonance position and energy. For these calculations, we take the beam and resonance parameters to be those of the present data-set: reaction ${}^{23}\mathrm{Na}(p,\gamma){}^{24}\mathrm{Mg}$, $E_0 = 11.964(17)$~MeV, $\epsilon = 5.118(7)\times 10^{-3}$~MeV/mm, $E_{cm} = 0.4906(3)$~MeV. This results in an expected resonance position of $Z_r=-7.04$~mm and expected resonance time of $T_r=-0.712$~ns. The calculations are performed only for the ``T'' method results from Section~\ref{sec:Bootstrap}, as these were shown to have good coverage properties for small sample sizes.

\begin{figure*}
\centering
\centerline{\includegraphics[width=140mm]{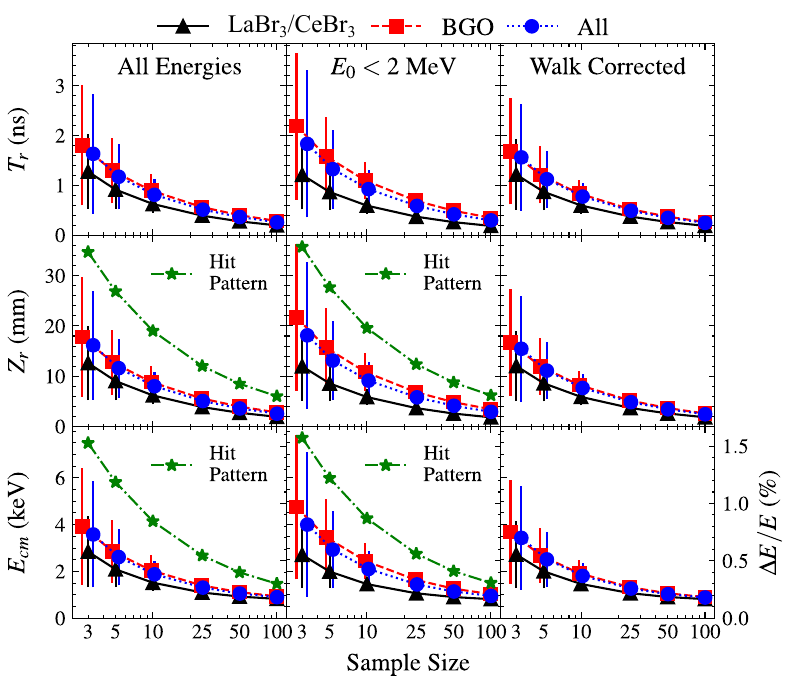}}
\caption{
\edit{
	Estimated confidence interval on the resonance time ($T_r$), position ($Z_r$), and center-of-mass resonance energy ($E_{cm}$) versus sample size ($N$) in the bootstrap pseudo-experiments. For each plot, the symbols indicate the mean confidence interval on the indicated parameter, across all bootstrap samples, and the error bars indicate the standard deviation of the confidence interval across all samples. In the bottom panels, the ticks on the left-hand side of the both plot indicate absolute confidence intervals on $E_{cm}$, while the ticks on the right-hand side of both plots indicate the relative uncertainty. Results are shown separately for the \labr{}/\cebr{} detectors, BGO detectors, and all detectors as indicated in the figure legend.
	 The left and center columns of the figure show results for samples taken at all $\gamma$-ray energies (left) and only when $E_0 < 2$~MeV (center). The rightmost column shows results including all energies, with the walk correction described in the text. All confidence intervals in this plot were obtained using the ``T'' method as explained in Section~\ref{sec:Bootstrap}. In the middle and lower rows (except for in the walk-corrected column), results of the traditional hit-pattern method of determining the $z$-centroid are shown as the green stars; see text for details.
}
}
\label{fig:bs_ez}
\end{figure*}

The results of the resonance position and energy confidence interval calculations are summarized in Table~\ref{tab:BootstrapEZ}. The table reports the mean and standard deviation of the confidence intervals on resonance position and energy obtained in each of the pseudo-experiments.  
\edit{The table shows results including events with all $\gamma$-ray energies, events with $E_0<2$~MeV, and all events including the walk correction described in the previous section. }
These results (along with the mean and standard deviation of the confidence intervals on $T_r$) are also plotted in Figure~\ref{fig:bs_ez}. For the smallest sample sizes, we find confidence intervals on the order of $2$--$4$~keV (relative confidence intervals ${\sim}0.4$--$0.8\%$), depending on the various detector and energy combinations. These decrease to below 1 keV ($<0.2\%$) for large sample sizes. Even the low-statistics results compare favorably with the ${\sim}0.5\%$ uncertainty obtainable with the hit-pattern method \cite{Hutcheon2012}, and the estimated uncertainties are well within the target of $\sigma_{E_{cm}} \lesssim 1\%$ needed for calculation of astrophysical reaction rates.

\edit{We emphasize that the results presented above only pertain to statistical uncertainties. 
One possible source of systematic uncertainty is the timing offset correction. Previously, we estimated that ignoring any energy-time correlations in the data used to determine timing offsets could result in systematic uncertainties as large as 0.65, 1.4, and 0.7~ns for the \labr{}/\cebr{}, BGO, and all detectors, respectively. These translate into respective systematic uncertainties on the resonance energy of $1.6$~keV ($0.32\%$), 3.1~keV ($0.62\%$), and 1.7~keV ($0.34\%$).
This systematic can be significantly reduced or eliminated by performing a walk correction such as the one employed here, or by selecting a sub-set of the calibration data whose $\gamma$-ray energies closely match those of the resonance of interest.
Another potential source of systematic uncertainty is potential shifts in the detector-by-detector timing offsets. We have estimated the magnitude of this systematic by comparing the timing offsets determined with the ${}^{23}\mathrm{Na}(p,\gamma){}^{24}\mathrm{Mg}$ resonant capture data with the synchronization offsets determined with the \nuc{60}{Co} source, which were taken over one month apart. A histogram of the timing offset shifts is well described by a Gaussian distribution with $\sigma = 2.68$~ns. Assuming the offset shifts are distributed randomly, and assuming equal numbers of detected events in each detector, this results in a systematic uncertainty on the mean TOF of $2.68~\mathrm{ns}/\sqrt{N_{det}}$, where $N_{det}$ is the number of detectors in the array. For the present $N_{det} = 20$ (\labr{}/\cebr{} detectors only), the systematic uncertainty on the mean TOF is $0.60$~ns. This translates to an uncertainty of $1.46$~keV on the resonance energy $(\Delta E/E=0.30\%)$. The impact of this systematic can potentially be reduced or eliminated by performing regular timing calibration runs throughout the measurement period.
}


Table~\ref{tab:BootstrapEZ} and Figure~\ref{fig:bs_ez} also present estimated confidence intervals on the resonance position and energy, using the traditional BGO hit pattern technique. These results were obtained by running a $10,000$ event GEANT3 simulation of the ${}^{23}\mathrm{Na}(p,\gamma){}^{24}\mathrm{Mg}$ reaction, with the standard BGO detector setup deployed in the simulation \cite{darioThesis}. The resulting standard deviation of the simulated BGO $z$~positions, \edit{without a cut on the BGO energy,} was $s_{G3} = 60.1(4)$~mm. This result was then used to calculate the estimated confidence intervals on the position, for sample size $N$, as $s_{G3}/\sqrt{N}$, and from there the confidence intervals on the resonance energy. These results are labeled ``Hit Pattern'' in both the table and the figure. In the table, the values in parenthesis represent the statistical uncertainties from the Monte Carlo statistics; the corresponding error bars are also present in the figure but are smaller than the data markers. \edit{The hit-pattern analysis was also repeated including the $E_0<2$~MeV cut on BGO energies ($2^{nd}$ column in both Table~\ref{tab:BootstrapEZ} and Figure~\ref{fig:bs_ez}). The impact of the energy cut is minor as it only impacts the analysis as a higher-order effect related to changes in the detection efficiency as a function of BGO $z$ position.}
%
%
The simulation results demonstrate that the resonance timing method significantly outperforms the hit-pattern method, especially for small sample sizes. At $N=3$, hit-pattern method is only able to determine the resonance position with an uncertainty of around $\nicefrac{1}{3}$ of the effective target length, and the goal of ${\sim}{1}\%$ uncertainty on $E_{cm}$ is only obtained for $N \sim 10$. The sensitivity of the two methods begins to converge at around $N=100$ events, and beyond this point the advantages of the resonance timing method are not likely to outweigh the additional experimental complexities.

\begin{table*}
\begin{center}
\caption{\edit{
	Confidence intervals on the mean resonance position ($\sigma_{Z_r}$) and center-of-mass resonance energy ($\sigma_{{E}_{cm}}$) extracted from the bootstrap pseudo-experiments. In each column, the leading value represents the average confidence interval across all bootstrap iterations, and the value in parenthesis represents the standard deviation of the confidence intervals. $N$ represents the number of samples of each iteration. As labeled in the table, the results are sub-divided into samples covering all $\gamma$-ray energies, samples with $E_0<2$~MeV, and all energies including a walk correction as described in the text. The results are further divided into samples from only the \labr{}/\cebr{} detectors; the BGO detectors; and all detectors. All results are calculated from $\sigma_{T_r}$ values obtained with the ``T'' method outlined in Section~\ref{sec:Bootstrap}. The portion of the table labeled ``Hit Pattern'' shows the estimated confidence intervals on resonance position and energy using the traditional hit-pattern technique applied to GEANT3 simulation results; see text for details.
}}
\label{tab:BootstrapEZ}
\begin{tabular}{c|cc|cc|cc}
\cline{2-7}
\multicolumn{1}{c}{} & \multicolumn{2}{c|}{All Energies} & \multicolumn{2}{c}{$E_0<2$ MeV} & \multicolumn{2}{c}{Walk Corrected} \\
\hline
$N$ & $\sigma_{Z_r}$ (mm) & $\sigma_{E_{cm}}$ (keV) &$\sigma_{Z_r}$ (mm) & $\sigma_{E_{cm}}$ (keV)&$\sigma_{Z_r}$ (mm) & $\sigma_{E_{cm}}$ (keV) \\\hline
\multicolumn{7}{c}{\labr{}/\cebr{}}\\
\hline
  3 & 12.6(7.4) & 2.85(1.51) & 12.0(6.9) & 2.71(1.40) & 12.0(7.0) & 2.72(1.43)\\
  5 & 9.03(3.81) & 2.09(76) & 8.56(3.50) & 1.99(69) & 8.60(3.61) & 2.00(72)\\
 10 & 6.18(1.77) & 1.52(33) & 5.85(1.59) & 1.46(29) & 5.88(1.68) & 1.46(31)\\
 25 & 3.86(68) & 1.10(11) & 3.65(60) & 1.06(9) & 3.68(65) & 1.07(10)\\
 50 & 2.72(34) & 0.924(46) & 2.57(29) & 0.904(38) & 2.59(32) & 0.907(42)\\
100 & 1.92(17) & 0.825(18) & 1.82(14) & 0.814(15) & 1.83(16) & 0.816(16)\\
\hline
\multicolumn{7}{c}{BGO}\\
\hline
  3 & 17.8(11.9) & 3.93(2.51) & 21.7(14.6) & 4.75(3.09) & 16.7(10.6) & 3.69(2.22)\\
  5 & 12.8(6.5) & 2.86(1.35) & 15.7(7.9) & 3.46(1.64) & 12.0(5.7) & 2.69(1.17)\\
 10 & 8.85(3.23) & 2.04(65) & 10.8(3.8) & 2.45(77) & 8.23(2.75) & 1.92(55)\\
 25 & 5.59(1.32) & 1.40(25) & 6.85(1.47) & 1.64(28) & 5.18(1.12) & 1.33(20)\\
 50 & 3.96(67) & 1.11(11) & 4.85(73) & 1.27(13) & 3.66(57) & 1.07(9)\\
100 & 2.81(34) & 0.936(47) & 3.43(36) & 1.03(6) & 2.59(29) & 0.907(38)\\
\hline
\multicolumn{7}{c}{All Detectors}\\
\hline
  3 & 16.2(10.8) & 3.58(2.27) & 18.1(13.0) & 4.00(2.74) & 15.5(10.2) & 3.44(2.13)\\
  5 & 11.6(5.9) & 2.62(1.21) & 13.1(7.2) & 2.93(1.49) & 11.1(5.5) & 2.52(1.14)\\
 10 & 8.01(2.92) & 1.88(58) & 9.13(3.54) & 2.10(71) & 7.66(2.72) & 1.81(54)\\
 25 & 5.06(1.21) & 1.31(22) & 5.80(1.43) & 1.44(27) & 4.83(1.11) & 1.27(20)\\
 50 & 3.58(62) & 1.05(10) & 4.12(72) & 1.14(12) & 3.42(56) & 1.03(9)\\
100 & 2.54(31) & 0.900(41) & 2.92(36) & 0.952(51) & 2.42(28) & 0.885(36)\\
\hline
\multicolumn{7}{c}{Hit Pattern}\\
\hline
3   & 34.7(2) & 7.49(5) & 35.7(6) & 7.71(12)&--&-- \\
5   & 26.9(2) & 5.82(4) & 27.7(4) & 5.99(9)&--&-- \\
10  & 19.0(1) & 4.15(3) & 19.6(3) & 4.27(6)&--&-- \\
25  & 12.0(1) & 2.68(2) & 12.4(2) & 2.75(4)&--&-- \\
50  & 8.50(6) & 1.96(1) & 8.75(14) & 2.01(3)&--&-- \\
100 & 6.01(4) & 1.47(1) & 6.19(10) & 1.51(2)&--&-- \\
%
\hline
\end{tabular}
\end{center}
\end{table*}

\section{Summary and Conclusions}
We have installed an array of 9 \labr{}, 11 \cebr{}, and 10 BGO $\gamma$-ray detectors surrounding the target of the DRAGON recoil separator and \edit{performed a first in-beam demonstration of} the array \edit{through} a study of the $E_{cm} = 0.4906(3)$~MeV resonance in the ${}^{23}\mathrm{Na}(p,\gamma){}^{24}\mathrm{Mg}$ reaction. For the sub-set of \labr{} and \cebr{} detectors, we performed coincidence timing measurements using a \nuc{60}{Co} source and found a $\gamma$-$\gamma$ timing spread best described by a Gaussian mixture distribution with total FWHM 0.443(2)~ns. For the \nuc{23}{Na} beam data, we analyzed the synchronized and calibrated RF TOF spectrum, in coincidence with \nuc{24}{Mg} recoils, and found that the time spread of the \labr{}/\cebr{} detectors is well described by a Gaussian distribution, with FWHM 4.57(26) ns. The same RF TOF spectrum for the BGO detectors is best described by a Gaussian mixture distribution, with FWHM 5.05(38)~ns. The RF TOF spectrum for all detectors is also well described by a Gaussian mixture distribution, with FWHM 4.67(31)~ns.

Using the measured RF TOF spectra, we performed a series of bootstrap pseudo-experiments to estimate the performance of the array in experiments measuring small numbers of recoil-$\gamma$ coincidences. For each pseudo-experiment, we estimated the sizes of the confidence intervals extracted for the resonance time, position, and energy. For small sample sizes, we found that the optimal method for estimating confidence intervals on the resonance time is to assume that the sample mean is distributed according to $\mathcal{T}_{N-1} s_{T_r}/N$, where $s_{T_r}$ is the standard deviation of measured times, $N$ is the number of detected events, and $\mathcal{T}_{N-1}$ is the Student's T distribution with $N-1$ degrees of freedom. This method gives the expected coverage properties for the confidence intervals, namely $1\sigma$ confidence intervals enclose the true value close to the expected $68.3\%$ of the time. Using this method, we found that expected (statistical) confidence intervals on the resonance time are on the order of 1--3~ns for $N=3$ events, decreasing to ${\sim} 0.2$--$0.3$~ns for $N=100$. The confidence intervals on position range from ${\sim} 12$--$20$~mm for $N=3$ down to ${\sim}1.5$--$4$~mm for $N=100$. Confidence intervals on the center-of-mass resonance energy range from ${\sim}1$--$4$ keV at $N=3$ down to ${\sim}0.8$--$1$~keV at $N=100$. A possible significant source of systematic uncertainty is the energy-time dependence of the timing offsets, which if left uncorrected could be as large as ${\sim}0.65$--$1.4$~ns in time, or ${\sim}1.5$--$3$ keV in energy. This systematic can be reduced or eliminated through proper treatment of energy-time correlations when synchronizing timing signals.

Overall, the present results are promising and indicate that the resonance timing method out-performs the hit-pattern method for determining radiative capture resonance energies with low sample sizes, and it is able to determine resonance energies with uncertainties below the $1\%$ level even with as few as $3$ detected events. This is particularly true for the \labr{}/\cebr{} detectors, but we also found promising results using only the BGO detectors.
One possible improvement to the present setup is to construct a dedicated, large-area array using a scintillator material with comparable time resolution to \labr{}/\cebr{}, as well as comparable detection efficiency to BGO. An attractive option is Lutetium-yttrium oxyorthosilicate (LYSO), which has the same density as BGO ($7.1$~g/cm$^3$) and decay times on par with \labr{} or \cebr{} (${\sim}30$~ns). LYSO also has reasonable energy resolution in between that of \labr{}/\cebr{} and BGO, e.g.\ $\Delta E/0.662~\mathrm{MeV} = 8.1\%$ FWHM \cite{Phunpueok2012}.

The resonance timing method represents a promising development for precisely determining the energies of astrophysical capture resonances, especially in low-yield experiments. \edit{This new technique is potentially applicable at DRAGON in future radioactive- and stable-beam experiments, as well as at other recoil separator facilities worldwide.}


\section*{Acknowledgments}
The authors are grateful to the staff at TRIUMF for their beam-delivery efforts throughout the experiment.
TRIUMF's core operations are supported via a contribution from the federal government through the National Research Council of Canada, and the Government of British Columbia provides building capital funds.
GC, ACE, ED, and JO acknowledge the support of the Natural Sciences and Engineering Research Council of Canada (NSERC), awards SAPIN-2020-00052 and RGPAS-2020-00001. DRAGON is generously funded through NSERC award SAPPJ-2023-00039.
PHR, ZP and SP were supported by the UK STFC UK Nuclear Data Network and the UK Science and Technology Facilities Council (STFC) via Grant Numbers ST/V001108/1 and ST/Y000358/1.
PHR, RS and SMC acknowledge support from the the UK National Measurements System Programmes Unit of the UK’s Department for Science, Innovation and Technology (DESIT).
UG acknowledges that this material is based upon work supported by the U.S. DOE, Office of Science, Office of Nuclear Physics under contract DE-FG02-93ER40789.


\end{document}